\documentclass[twocolumn,prl,aps,floatfix]{revtex4}% Physical Review B

\usepackage{graphicx}% Include figure files
\usepackage{dcolumn}% Align table columns on decimal point
\usepackage{bm}% bold math

\begin{document}

\preprint{APS/123-QED}

\title{Nernst effect of  epitaxial
Y$_{0.95}$Ca$_{0.05}$Ba$_{2}$(Cu$_{1-x}$Zn$_{x}$)$_{3}$O$_{y}$ and
Y$_{0.9}$Ca$_{0.1}$Ba$_{2}$Cu$_{3}$O$_{y}$ films}

\author{I. Kokanovi\'{c}$^{1,2}$,
J.R. Cooper$^1$ and M. Matusiak$^{1}$},

\affiliation{$^1$Department of Physics, University of Cambridge, J. J. Thomson Avenue,
Cambridge CB3 OHE, U.K.\\
$^2$Department of Physics, Faculty of Science, University of Zagreb, P.O.Box 331,
Zagreb, Croatia.}
 \altaffiliation{Electronic address: kivan@phy.hr}

 %Lines break automatically or can be forced with \\
%\author{}%

%\affiliation{%
%\\IRC in Superconductivity and
%Department of Physics, University of Cambridge, Madingley Road, Cambridge
%CB3 OHE, U.K.
%}%

\date{\today}% It is always \today, today,
             %  but any date may be explicitly specified

 %Valid PACS numbers may be entered using the \verb+\pacs{#1}+  command.
 \begin{abstract}
 We report Nernst effect measurements of some crystalline films grown
by pulsed laser deposition, namely  slightly under- and nearly optimally-doped
Y$_{0.95}$Ca$_{0.05}$Ba$_{2}$(Cu$_{1-x}$ Zn$_{x}$)$_{3}$O$_{y}$ (with $x$ = 0, 0.02
and 0.04) and over-doped Y$_{0.9}$Ca$_{0.1}$Ba$_{2}$Cu$_{3}$O$_{y}$. We argue that
 our results and most of the data for LSCO \cite{Xu} are consistent with the theory of
Gaussian  superconducting fluctuations \cite{Ussishkin}.

 \end{abstract}

% PACS, the Physics and Astronomy
                             % Classification Scheme.
%\keywords{Suggested keywords}%Use showkeys class option if keyword
                              %display desired
\maketitle

In conventional type II  superconductors, the motion of Abrikosov vortices induced by
a thermal gradient ($\nabla_xT$), perpendicular to the magnetic field $B$, gives rise
to a transverse electric field $E_y$ and hence a  Nernst voltage, the Nernst
coefficient $\nu$ being defined by the relation, $\nu=\frac{E_y}{\nabla_xTB}$. In some
influential papers, measurements of significant Nernst signals  over a
 broad temperature range  \emph{well above} the superconducting transition temperature ($T_c$)
 have
 been reported, initially for
 La$_{2-x}$Sr$_x$CuO$_4$ (LSCO) \cite{Xu} and later for several other cuprate crystals \cite{Wang06}.
 These results have been interpreted  as evidence for the
  existence of vortex-like excitations above $T_c$,  and for two separate
  temperature scales for phase and amplitude fluctuations of the superconducting order
   parameter. Because $\nu$
 seemed to be particularly large  for under-doped samples,
 in the pseudogap region of the cuprate phase diagram, it was suggested
  that the pseudogap is actually caused by  superconducting fluctuations, in contradiction to
  arguments based on heat capacity studies \cite{Loram1}.
   More recently, by  introducing controlled
  amounts
  of disorder by electron irradiation \cite{Rullier} or by Zn doping \cite{XuZn} other
  authors have shown that the onset of  a larger Nernst signal is not linked to $T^*$,
  the characteristic energy scale of the pseudogap.   The Nernst data \cite{Xu,Wang06} have also been cited
  by many authors in support
   of the scenario in which the pseudogap remains finite over the whole superconducting region of the cuprate phase diagram
   rather than going to zero for slightly over-doped samples.

Nernst effect studies of  the cuprates  inspired an extension of the theory of
superconducting fluctuations \cite{Larkin} by   Ussishkin et al. \cite{Ussishkin} who
showed that for weak (Gaussian) fluctuations (GF), the off-diagonal term,
$\alpha_{xy}^{s}$, of the Peltier tensor is given by:

 \begin{equation}
{\alpha_{xy}^{s} = \frac{k_Be}{3h}
\frac{\xi_{ab}^{2}}{l_B^{2}s}\frac{1}{\sqrt{1+(2\xi_{c}/s)^{2}}}}\label{1}
 \end{equation}
here $\xi_{ab}$ and $\xi_{c}$ are the temperature-dependent coherence lengths
 parallel  and perpendicular to the layers, $s$ is the interlayer spacing, $l_B=(\hbar/eB)^{1/2}$
 is the magnetic length and the anisotropy $\gamma$=$\xi_{ab}$/$\xi_{c}$. The fluctuation contribution to the
 Nernst coefficient  is given by:

  \begin{equation}
{\nu_{s}= \alpha_{xy}^{s}/[\sigma(T)B]}\label{2}
 \end{equation}
where $\sigma(T)$ is the total electrical conductivity. Ussishkin et al.
\cite{Ussishkin} found that
   Gaussian superconducting fluctuations  account well for  Nernst data of optimally doped and over-
   doped La$_{2-x}$Sr$_x$CuO$_4$ crystals
with $x$ = 0.20 and 0.17 but for   an under-doped sample with $x=0.12$ they suggested
that stronger non-Gaussian fluctuations give a larger Nernst signal and also reduce
$T_c$ from the  mean field value. Recently their GF theory \cite{Ussishkin} was
verified over a wide temperature range by experiments on thin amorphous
low-temperature superconductors \cite{Pourret}, this is especially important in view
of an alternative theoretical viewpoint reported recently \cite{Sergeev}.

     Here we report  measurements of the Nernst effect for the same Ca and Zn substituted YBa$_2$Cu$_3$O$_{6+x}$
      (YBCO) epitaxial films
      for which
     in-plane resistivity and  magnetoresistivity $\rho(B,T)$, and Hall coefficient $R_H$,
data were previously reported \cite{Kokanovic}.
 We show that
 our Nernst data above $T_c$ are consistent with GF theory
  \cite{Ussishkin}.
 Although  the Nernst signal is more clearly visible in our Zn-doped
 samples,  we argue that this is primarily  because of their smaller
conductivity. In  contrast to the suggestion of Refs. \onlinecite{Rullier} and
\onlinecite{XuZn}, for our samples there is no evidence for the Nernst signal being
enhanced by another mechanism such as inhomogeneous superconductivity. We also show
that GF   can account for the general behavior of   Nernst data of LSCO \cite{Xu} over
the whole doping range.

  Values of the hole concentration $p$ determined  from
the room-temperature thermopower, $S(290K)$ \cite{Obertelli}, are given in Table I
together with $T_c$ values and transition widths (FWHM, $\delta T_c$, in $d \rho
/dT$). Small changes in $p$ have occurred since the previous work and therefore
quantities such as $\rho(B,T)$ and $R_H$ were measured again below 120 K. In the
Nernst set-up used here the 10 x 5 x 1 mm$^3$ SrTiO$_3$  substrate was glued between
copper and stainless steel posts each holding a heater and a small Cernox thermometer.
A sketch of the patterned thin film is shown in the insert to Fig. 1(a). The
temperature gradient of 2 or 4 K/cm was applied along the longitudinal direction, $B$
was applied along the c direction of the film (perpendicular to the surface of the
substrate) and the Nernst
            signal was measured     between the ``Hall contacts''  using a Keithley Model 182
nanovoltmeter. The transverse voltage $V_B$ was measured for $+B$ and $-B$ while
sweeping either  $T$ or $B$,  the Nernst voltage was defined as
$\frac{1}{2}(V_B-V_{-B})$ and converted to electric field using the distance (1.5 mm)
between the inside edges of opposite gold contact pads. $\nabla T$ was checked by
measuring the thermoelectric voltage between two longitudinal contacts. The precise
temperature of the sample was determined by comparing $\rho(T ,B)$ data measured with
and without an applied temperature gradient.

\begin{figure}
\includegraphics[width=7.5cm,keepaspectratio=true]{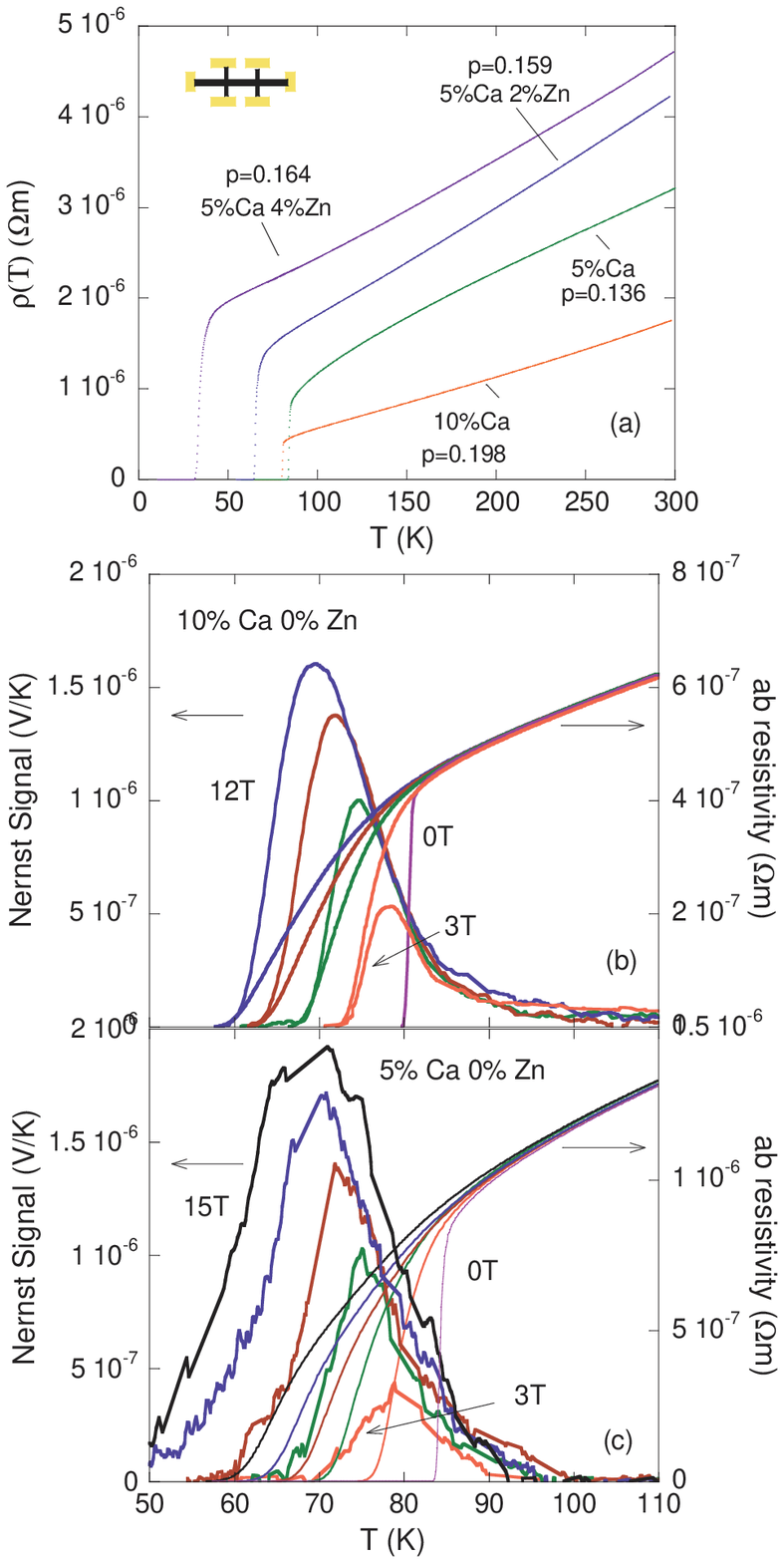}
\caption{ Color online. (a) In-plane resistivity versus temperature for
Y$_{0.95}$Ca$_{0.05}$Ba$_2$(Cu$_{1-x}$Zn$_x$)$_3$O$_y$ ($x$ = 0, 0.02 and 0.04) and
Y$_{0.9}$Ca$_{0.1}$Ba$_{2}$Cu$_3$O$_7$ films. (b) Nernst signal versus $T$ at 3, 6, 9
and 12 T for the Y$_{0.9}$Ca$_{0.1}$Ba$_{2}$Cu$_3$O$_y$  film. (c) Nernst signal
versus $T$ at 3, 6, 9, 12 and 15 T for the Y$_{0.95}$Ca$_{0.05}$Ba$_{2}$Cu$_3$O$_y$
film. In (b) and (c) in-plane resistivities at the same fields and at 0 T are also
shown. }
 \label{rawdata1}
\end{figure}

Zero-field $\rho(T)$ data for the four films are shown in Fig. 1(a). Representative
$\rho(T,B)$ and Nernst  data are shown for the  over-doped 10 $\%$ Ca sample ($p=
0.198$) in Fig. 1(b) and for the most under-doped 5 $\%$ Ca sample ($p= 0.136$) in
Fig. 1(c). The $\rho(T,B)$ curves show the usual ``fanning out'' property which is
typical of the cuprates but is  not observed in conventional type II superconductors.
The points at which $\rho(T,B) \simeq 0$ on the scales shown correspond to the
irreversibility line $B_{irr}(T)$. Many researchers consider that  for $B >
B_{irr}(T)$ there  is a wide ``vortex liquid'' region where
 vortices are still present but
 no longer form a regular lattice  and are no longer
pinned. We have argued previously \cite{Cooper,CooperBabic} for an
alternative viewpoint in which the vortices disappear for $B$
equal to, or slightly greater than, $B_{irr}(T)$. In other words
 $B_{irr}(T)$ could actually be the $B_{c2}(T)$ line which has
been heavily suppressed by superconducting fluctuations that may be further enhanced
by the magnetic field. This view is still controversial but is not inconsistent with a
recent dynamical scaling analysis of voltage-current measurements for  YBCO single
crystals and films \cite{Lobb}.

\begin{figure}
\includegraphics[width=8.5cm,keepaspectratio=true]{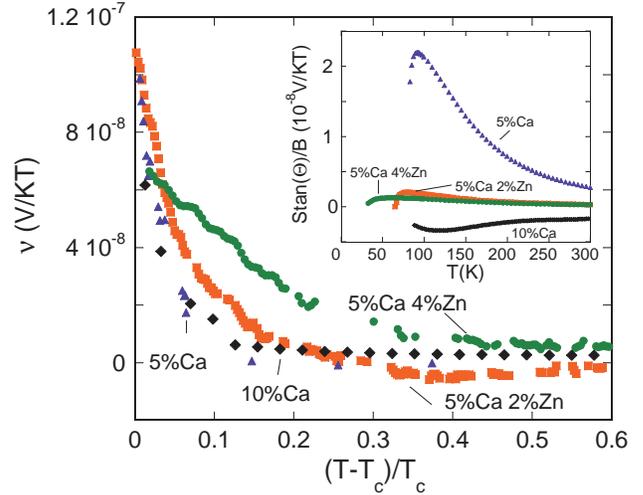}
\caption{ Color online: Nernst coefficient, $\nu$, versus reduced temperature $t$ for
the Y$_{1-z}$Ca$_{z}$Ba$_2$(Cu$_{1-x}$Zn$_x$)$_3$O$_y$ ($z$ = 0.05, 0.1; $x$ = 0, 0.02
and 0.04) films. Inset shows $S \tan(\theta)/B$, for the same samples, where $S$ is
the thermopower and $\theta$ the Hall angle in a field  $B=6 T$. }
\end{figure}

 If one does assume that vortices are still present well above
  $B_{irr}(T)$ then  the ratio
 $\nu(T,B)/\rho(T,B)$ can be used to determine the entropy per vortex as has been done
 for LSCO \cite{Capan}. This assumes
 isotropic vortex pinning forces, since the resistivity arises
 from sideways motion of the vortices (perpendicular to the direction of current flow) while
  the  Nernst
 voltage  arises
 from the flow of vortices along the length of the sample.  In Fig. 1(b) the
  onset of the Nernst signal is the same as the onset of
 resistivity to within experimental uncertainty of $\pm$ 0.5 K while data for the
 under-doped 5 $\%$ Ca sample in Fig. 1(c)  show  sizeable Nernst
 signals
 $\sim$ 6 K below the points at which $\rho(T,B)= 0$. Vortex  pinning at
  twin boundaries
is known to be important  in YBCO based compounds and can be highly anisotropic
 \cite{Fleshler}. We therefore believe that the different behavior of the two samples
is not directly linked to their different doping levels,  but arises because
anisotropic pinning by twin boundaries is significant for the data in Fig. 1(c) but
not for those in Fig. 1(b).

\begin{figure}
\includegraphics[width=8.5cm,keepaspectratio=true]{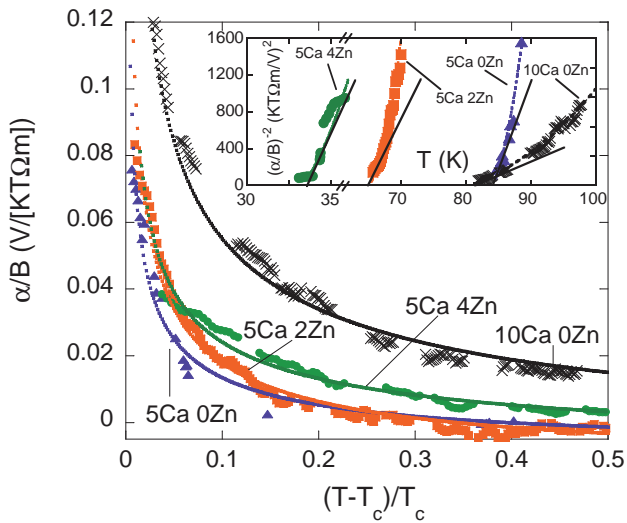}
\caption{ Color online: $ \alpha/B \equiv \sigma_{ab}(T) \nu$ versus $t$ for
Y$_{0.95}$Ca$_{0.05}$Ba$_{2}$(Cu$_{1-x}$Zn$_x$)$_3$O$_y$ ($x = 0$, 0.02, 0.04)
 and Y$_{0.9}$Ca$_{0.1}$Ba$_{2}$Cu$_{3}$O$_{y}$ films. The thin
dotted lines show fits to Eq. 1. The inserts are plots of $(\alpha /B)^{-2}$ versus
$T$ near $T_c$, with dashed lines showing the fits to Eq. 1 and solid lines the linear
3D limit of Eq. 1. }
\end{figure}

Figure 2 shows the Nernst coefficient above $T_c$ for all four samples \textit{vs.}
the reduced temperature $t
  \equiv(T-T_c)/T_c$ up to $t$ = 0.6. Here the initial linear part of the Nernst
  voltage versus $B$ curve has been used to determine $\nu$.  As $T_c$ is approached from above, these
  curves  become  non-linear   at a field of
   1-2 x $(T-T_c)$ Tesla.
This is to be expected for GF which are gradually suppressed when
 $l_B$ becomes comparable with $\xi_{ab}(T)$, or equivalently when
   $B\sim (T-T_c)\mid dB_{c2}/dT\mid$ where
   $dB_{c2}/dT$ is the slope of $B_{c2}$ just below $T_c$ and is  1-2 T/K for cuprates with $T_c$ values of 80 to 90 K.
   The criterion used in
  Ref. \onlinecite{Xu} for a significant ``vortex'' signal is $\nu$ = 4
  nV/K-Tesla. At first sight this  might suggest that  there are vortices up to $t \simeq 0.6$
in our
  4 $\%$ Zn doped sample, and that disorder increases the  temperature difference
 between the  formation of
  fluctuating Cooper pairs and the onset of
 phase coherence, as proposed for electron-irradiated
   YBCO samples \cite{Rullier}. However we reject this hypothesis and  argue below that
    the observed value of $\nu$ arises
   simply from GF.
  One reason for the apparent enhancement of $\nu$ is that any normal
  state (quasiparticle)  contribution $\nu_n$ is suppressed by Zn doping. $\mid\nu_n\mid$ is expected to be
   smaller than $\mid S \tan \theta_H\mid$ where
  $S$ is the  thermoelectric power and $\theta_H$ the Hall angle in the normal
  state, given by
  $\tan \theta_H\equiv\sigma_{xy}/\sigma_{xx}=\rho_{xy}/\rho_{xx}$. The condition
 $\mid\nu_n \mid \ll\mid S\tan \theta_H\mid$ arises from the Sondheimer
  cancellation \cite{Sond} between the off-diagonal thermal and electrical
  currents that is exact for a single parabolic band with an energy
  independent relaxation time \cite{Wang01}.  If these rather
  restrictive conditions do not apply, then we expect
  that $\mid\nu_n\mid\sim\mid S\tan\theta_H\mid$. As shown in the
  inset to Fig. 2, for the two Zn doped films, $S \tan
  \theta_H$ is particularly small which makes the GF  term more
  visible.

  Fig.  3 shows $\alpha/B \equiv\sigma_{ab}(T) \nu$ vs. $t$ for
  the same samples as in Fig.  2 together with fits
  to Eq. 1 with $s$ = 1.17 nm, the $c$-axis lattice parameter.   An extra
    fitting parameter, a small offset $\simeq$ -0.01 V/K-T-$\Omega$m,
    has been included in Eq. 1 to account for $\nu_n$.
  In the 3D limit of Eq. 1 near $T_c$ where $\xi_{c}(T)\gg s$, we expect
   $\alpha^{-2} \propto
  (T-T_c)$. Corresponding plots are shown in the insert to Fig. 3.  There
  are small linear regions  extrapolating to $y=0$  near the measured
  value of $T_c$. At higher $T$ these cross over to the quadratic law,  $\alpha^{-2} \propto
  (T-T_c)^2$   expected in the 2D limit.   A similar
   analysis  \cite{LoramCooper} of the heat capacity of several cuprate families
showed that the difference, $\Delta T_c$, between the measured value, $T_c^m$, and the
fitted or linearly extrapolated value, $T_c^f$, was caused  by strong (critical)
fluctuations. $\Delta T_c$ was $\sim$ 1 K for YBCO samples, as found for the present
Nernst data, and $\sim$ 5 K for other extremely anisotropic cuprates.

The fitting
  parameters $ \xi_{ab}(T=0)$ and
    $ \gamma \equiv \xi_{ab}/ \xi_{c}$ are summarized in Table I.
 The value $\xi_{ab}(0)$ = 1.6 nm for
  the 5 $\%$ Ca sample corresponds to $B_{c2}(0) \equiv
  \Phi_0/2\pi\xi_{ab}(0)^2$ =130 T for $B \parallel c$ and also
  agrees with the value obtained  by GF analysis of
  heat capacity data  \cite{LoramCooper}. The  value of $\gamma$
  also agrees with other estimates for well-oxygenated YBCO \cite{CooperBabic}.
  The values of $\xi_{ab}(0)$  for the two Zn-doped films are
  larger. For the 2 $\%$ Zn film,
  $1/\xi_{ab}(0)$ scales with $T_c$ as expected, but for the 4 $\%$ Zn
  film
  the short mean free path probably reduces
  $\xi_{ab}(0)$ according to the standard
  dirty-limit formula \cite{Waldram}.  The coherence length of the
   10 $\%$ Ca,  0 $\%$ Zn film is  longer than that for the
  5 $\%$ Ca,  0 $\%$ Zn film. This is not understood, however $\rho(T)$ is a factor of 2
  smaller, and also
   for  10 $\%$ Ca,  Eq. 1 gives a good fit with
     $\xi_{ab}(0)$ = 2.2 nm and  $\gamma$ = 12
    over a smaller range of $t$ (between 0.03 and 0.2).

\begin{figure}
\includegraphics[width=8.5cm,keepaspectratio=true]{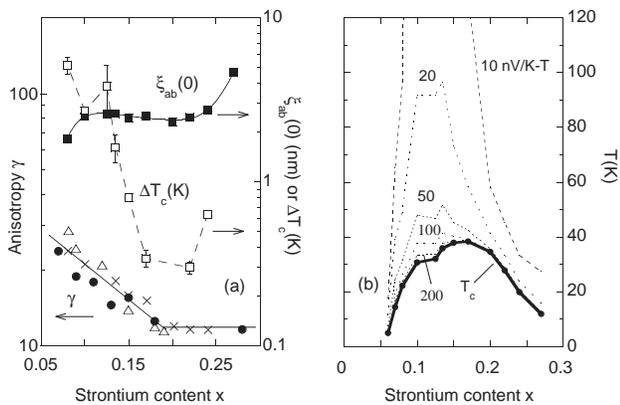}
\caption{(a) Left hand scale, anisotropy, $\gamma \equiv \xi_{ab}(0)/ \xi_{c}(0)$,
obtained from the room temperature resistivity anisotropy \cite{Kimura} ($\bullet$)
and two sets of London penetration depth data at low $T$ ($\times$)
  \cite{Panagop} and ($\triangle$) \cite{Maeda}, vs Sr content $x$ in LSCO. Straight lines
show  average values of $\gamma$ used in calculations.  Right hand  scale,
$\xi_{ab}(0)$ obtained from a GF
 analysis \cite{LoramCooper,Cvfootnote} of heat capacity (C$_v$) data \cite{Loram1}
above $T_c$  using the same values of $\gamma$.   $\triangle T_c(x)$ \cite{Cvfootnote}
is related to the  strength of critical (non-Gaussian) fluctuations
\cite{LoramCooper}. (b) Calculated constant $\nu$ contours in the $(T,x)$ plane, using
Eqs. 1 and 2, with $s$ = 0.66 nm, $\rho_{ab}(x,T)$ from Ref. \onlinecite{Ando} and
$\xi_{ab}(0)$ and $\gamma$ from Fig. 4(a). }
\end{figure}

\begin{table}
\caption{Summary of results}
\begin{ruledtabular}
\begin{tabular}{l|c|c|c|c|c}
$Sample$&$T_c $&$\delta T_c$&$p $&$\xi_{ab} $&$\gamma$\\
$Ca,Zn$ &$ (K)$&$  (K)$&$(holes/Cu)$&$(nm)$&$ $\\
\hline $0.05$ &$84.2$&$0.6$ &0.136 $\pm0.002$&1.6$\pm0.2$&6.2$\pm0.5$\\
\hline $0.05, 0.02$ &$65.1$&$1$ &0.159 $\pm0.004$&1.9$\pm0.2$&7.2$\pm0.5$\\
\hline $0.05, 0.04$ &$33.3$&$1.5 $&0.164 $\pm0.004$&2.6$\pm0.2$&5.1$\pm0.5$\\
\hline $0.1$ &$80.6$&$0.7$&$0.198\pm0.004$&3.4$\pm0.2$&7.5$\pm0.5$\\
\end{tabular}
\end{ruledtabular}
\label{summary}
\end{table}

  The success of the GF analysis described above
  encouraged us to look again  at published data for LSCO crystals,
  since Fig. 4 of Ref. \cite{Xu} and other versions \cite{Wang06,Wang01}  provide key support for
  the alternative, widely accepted, phase fluctuation and pseudogap pictures.
  In Fig. 4(a) we show values of $\gamma$ obtained from the anisotropy in the London penetration depth
   at low $T$ \cite{Panagop,Maeda} and from that in $\rho(300 K)$  \cite{Kimura} as
   well as values of $\xi_{ab}(0)$ obtained from GF  analysis  \cite{LoramCooper} of
   the electronic specific heat \cite{Loram1} above $T_c$. These have been
   used in Eqs. 1 and 2, together with the measured values of $T_c(x)$ \cite{Loram1}
    and $\rho_{ab}(T,x)$  \cite{Kimura,Ando}, to
    calculate the
   contour plots for $\nu$
   shown in Fig. 4(b).  This   GF
  picture correctly accounts for the   peaked
  structure of $\nu$ vs. Sr content ($x$) and the  magnitude of
  $\nu$ between 40 and 80 K \cite{Xu,Wang06,Wang01}.  The asymmetric GF peak arises  from
  the dome-shaped
    $T_c(x)$ curve and the approximately linear increase of  $\sigma_{ab}(T)$
  with $x$ \cite{Kimura,Ando}.    Above 80 K  GF  theory gives values of
  $\nu$ which are too large,  possibly because the fluctuations are suppressed by inelastic scattering
  processes. In the experimental contour plot \cite{Xu}, there is a small region, $x\leq 0.13$
  and $T-T_c\leq 5$ K, where $\nu\simeq$ 500 nV/K-T is too large to be consistent with
  GF theory.  Fig. 4(a)
   shows that  $\Delta T_c \sim 2-5$ K for $x\leq 0.13$,  supporting the idea
  \cite{Ussishkin} that in this small region $\nu$ is enhanced by critical (non-Gaussian)
  fluctuations.

In summary we find that weak (Gaussian) superconducting fluctuations account for our
Nernst data for YBCO ab-plane films substituted with various levels of Ca and Zn,
until at least 30 K above $T_c$. They also account for the main features of the Nernst
contour plots for LSCO crystals.

We are grateful to J. W. Loram and S. H. Naqib for fruitful collaboration and many
useful suggestions. This work was supported by EPSRC (UK), grant number EP/C511778/1
and the Croatian Research Council, MZOS project No. 119-1191458-1008.


\begin{thebibliography}{99}
\bibitem{Cvfootnote} Data for the electronic heat capacity, $C_v$, of  LSCO \cite{Loram1}
 were fitted to the GF expression (Eq. 1 of Ref. \onlinecite{LoramCooper})
 plus a linear background. Here $\Delta T_c \equiv
T_{c}^{m} - T_{c}^{f}$, where $T_{c}^{m}$ is the midpoint of the step in $C_v/T$ and
$T_{c}{^f}$ is obtained from the GF fits.
\end{thebibliography}
\end{document}